\documentclass[conference]{IEEEtran}
\usepackage{hyperref}
\IEEEoverridecommandlockouts
\usepackage{cite}
\usepackage{amsmath,amssymb,amsfonts}
\usepackage{algorithmic}
\usepackage{graphicx}
\usepackage{textcomp}
\usepackage{csquotes}
\usepackage{xcolor}
\def\BibTeX{{\rm B\kern-.05em{\sc i\kern-.025em b}\kern-.08em
    T\kern-.1667em\lower.7ex\hbox{E}\kern-.125emX}}

\begin{document}

\title{User-Centric Evaluation Methods for Digital Twin Applications in Extended Reality\\
\thanks{{$\copyright$ \the\year~IEEE. Personal use of this material is permitted. Permission from IEEE must be obtained for all other uses, in any current or future media, including reprinting/republishing this material for advertising or promotional purposes, creating new collective works, for resale or redistribution to servers or lists, or reuse of any copyrighted component of this work in other works. This preprint has not undergone any post-submission improvements or corrections. To cite this article: F. Vona, M. Warsinke, T. Kojić, J. -N. Voigt-Antons and S. Möller, "User-Centric Evaluation Methods for Digital Twin Applications in Extended Reality," 2025 IEEE International Conference on Artificial Intelligence and eXtended and Virtual Reality (AIxVR), Lisbon, Portugal, 2025, pp. 142-146, doi: 10.1109/AIxVR63409.2025.00028.}}
}

%\author{\IEEEauthorblockN{Anonymous Authors}}

\author{\IEEEauthorblockN{1\textsuperscript{st} Francesco Vona}
\IEEEauthorblockA{\textit{Immersive Reality Lab} \\
\textit{Hochschule Hamm-Lippstadt}\\
Lippstadt, Germany }
\and
\IEEEauthorblockN{2\textsuperscript{nd} Maximilian Warsinke}
\IEEEauthorblockA{\textit{Quality and Usability Lab} \\
\textit{Technische Universität Berlin}\\
Berlin, Germany}
\and
\IEEEauthorblockN{3\textsuperscript{rd} Tanja Kojić}
\IEEEauthorblockA{\textit{Quality and Usability Lab} \\
\textit{Technische Universität Berlin}\\
Berlin, Germany}
\and
\IEEEauthorblockN{4\textsuperscript{th} Jan-Niklas Voigt-Antons}
\IEEEauthorblockA{\textit{Immersive Reality Lab} \\
\textit{Hochschule Hamm-Lippstadt}\\
Lippstadt, Germany}
\and
\IEEEauthorblockN{5\textsuperscript{th} Sebastian Möller}
\IEEEauthorblockA{\textit{Quality and Usability Lab} \\
\textit{Technische Universität Berlin and DFKI}\\
Berlin, Germany}
}

\maketitle

\begin{abstract}
The integration of Digital Twins (DTs) with Extended Reality (XR) technologies, such as Virtual Reality (VR) and Augmented Reality (AR), is transforming industries by enabling more immersive, interactive experiences and enhancing real-time decision-making. User-centered evaluations are crucial for aligning XR-enhanced DT systems with user expectations, enhancing acceptance and utility in real-world settings. This paper proposes a user-centric evaluation method for XR-enhanced DT applications to assess usability, cognitive load, and user experience. By employing a range of assessment tools, including questionnaires and observational studies across various use cases—such as virtual tourism, city planning, and industrial maintenance—this method provides a structured approach to capturing the user’s perspective.
\end{abstract}


\section{Introduction}

A Digital Twin (DT) is a virtual representation of a physical object or system that enables real-time monitoring, simulation, and optimization \cite{grieves2014}. DTs allow for real-time updates, scenario testing, and predictive maintenance, making them highly relevant for today’s fast-evolving industries, including manufacturing, urban development, and infrastructure management \cite{negri2017, tao2019, fuller2020}.

The emergence of Extended Reality (XR) technologies, which encompass Virtual Reality (VR), Augmented Reality (AR), and Mixed Reality (MR), has enhanced the usability and potential of DTs, adding an interactive, immersive dimension. VR offers an immersive environment in which users can explore DTs in 3D, enabling a more detailed visualization and analysis of complex data. For instance, VR-driven DTs are particularly valuable in training scenarios where users can interact with a realistic copy of a physical asset, practicing procedures without safety risks and gaining insights into potential operational challenges \cite{jones2020}.

AR further extends the capabilities of DTs by overlaying digital information onto the physical world, making it possible to access real-time data while interacting with physical assets. For example, a maintenance technician could use an AR interface to visualize a DT’s diagnostic information overlaid on machinery, allowing for precise adjustments and real-time feedback. This real-time overlay makes DTs more actionable and accessible for tasks that require quick decision-making, such as field inspections or on-site monitoring \cite{lu2020}.

These XR applications, including VR’s immersive simulations and AR’s real-time overlays, transform the interaction with DTs into a more intuitive and collaborative experience. This integration allows for enhanced data visualization, a deeper understanding of system behaviors, and more effective training and operational efficiency across industries.

However, there remains a significant lack of consensus on standardized methods for evaluating their effectiveness, particularly in XR-enhanced applications. Although various frameworks have been proposed to assess DT performance and impact, they frequently lack comprehensive benchmarks that account for the unique influence of XR on usability, engagement, and overall user experience (UX) \cite{qi2018}.  Recent efforts to address this gap have shown promise, yet the solutions remain immature and have not been widely accepted or standardized \cite{khadim2023uxdigitaltwins, manickam2023uxdtmm}. 

Traditional evaluation methods tend to emphasize functional advancements, such as monitoring, simulation, and autonomy. However, they often neglect the critical element of human interaction, where users actively engage with DTs for exploration, co-creation, and decision-making. By focusing on experiential factors, user-centric evaluations ensure that DTs are not only functionally robust but also intuitive, engaging, and even delightful for users. This human-centered perspective is key to fostering broader adoption and maximizing the real-world impact of XR-enhanced DTs \cite{liu2021, paelke2018}.

This article identifies and addresses the gap in evaluation methodologies for XR-enhanced DTs focusing on a systematic, user-centric approach. By integrating immersive and experiential factors into the evaluation framework, we aim to bridge the divide between technical functionality and human interaction.

\section{State of the art}
The integration of DT and XR technologies has emerged as a cornerstone of advancements in Industry 4.0 and 5.0, enabling applications across diverse sectors such as manufacturing, healthcare, and collaborative design \cite{tao2019digital,soori2023digital}. Recent studies have introduced a variety of methodologies and conceptual frameworks that leverage these technologies to optimize industrial systems, enhance productivity, and facilitate advanced human-machine interactions. For example, the application of a co-simulation environment combining DT with VR has been investigated to optimize cyber-physical production systems (CPPS) within industrial workstations, particularly focusing on ergonomic assessments and refining human-robot collaborative workspaces for enhanced safety and efficiency \cite{havard2019integration}. Such studies highlight DT’s role in simulating, monitoring, and enhancing CPPS functionality within the frameworks of Industry 4.0.

Further expanding DT capabilities, various frameworks target interoperability between DT descriptions and XR applications, enabling scalable XR solutions in smart manufacturing environments \cite{tu2023twinxr}. These solutions, tested in operational use cases like overhead cranes and robotic arms, improve real-time data transfer, facilitate seamless information exchange, and optimize task execution, underscoring the utility of XR-DT combinations in industrial applications. Similarly, XR-DT frameworks have been applied to streamline production alignment and optimize workforce allocation, as seen in resource-intensive contexts such as brake disc manufacturing, where optimal deployment of operators and automated guided vehicles is achieved through real-time data synchronization \cite{Catalano2022-sd}.

In healthcare, DT technologies are being explored for applications in personalized and predictive medicine, telemedicine, and VR-based medical education, all within the emerging metaverse framework \cite{Turab2023-af}. These DT-based solutions hold transformative potential, enhancing patient engagement, improving remote care capabilities, and boosting healthcare system efficiency. In manufacturing, DT-driven XR frameworks designed for industrial systems, such as crane operations, also facilitate human-machine interactions (HMI), synchronizing real-time data with physical counterparts to improve productivity, safety, and data-driven decision-making \cite{Yang2022-ll}.

Contributions to this field also include reviews highlighting DT’s application in smart manufacturing, where it enhances lifecycle monitoring, process optimization, and predictive maintenance. By leveraging DTs, manufacturers can streamline workflows, optimize supply chain management, and improve equipment reliability through continuous monitoring and real-time simulation \cite{tao2019digital,soori2023digital}. In collaborative design, DT and XR provide immersive environments conducive to value co-creation and real-time feedback on project progress, aligning with circular economy frameworks and long-term sustainability goals \cite{Bertoni2023-ad}. Additionally, interaction frameworks combining VR, AR, and MR with DT systems enable immersive environments that support complex industrial tasks, such as equipment maintenance and operator training, enhancing operational efficiency and user engagement \cite{Ke2019-mj}.

Despite these advancements, a notable limitation across this body of work is the lack of empirical evaluations with real users, leaving gaps in understanding how these technologies impact UX and engagement. User-centered evaluation methods, such as usability testing \cite{nielsen1993usability}, cognitive walkthroughs \cite{lewis1990cognitive}, and think-aloud protocols \cite{ericsson1980thinkaloud}, could provide valuable insights into the practical implementation of DT-XR systems. Usability testing, for instance, assesses the ease of interaction, error rates, and overall satisfaction of users in real-world scenarios, offering quantifiable data to optimize interface design. Cognitive walkthroughs involve domain experts who systematically explore the task sequences to identify potential usability issues, which can be particularly useful in identifying pain points in DT-XR applications before implementation in operational environments \cite{lewis1990cognitive}. Think-aloud protocols, where users verbalize their thought process while interacting with the system, offer insights into cognitive load and mental models, providing crucial feedback for enhancing the intuitiveness of DT-XR interfaces \cite{ericsson1980thinkaloud}.

Integrating these user-centric methodologies would offer a comprehensive perspective on user satisfaction, cognitive load, and ease of interaction, bridging the current gap between technical functionality and UX. Such empirical studies are essential for ensuring that DT-XR technologies are not only robust in their operational capacities but also aligned with user needs, enhancing both acceptance and overall efficacy. As these technologies continue to evolve, future work should prioritize user-centered research methodologies, contributing to the development of DT-XR solutions that are both technologically sound and designed with the end-user in mind.

\section{Methods}

\subsection{General Objective}
We outline the methods for validating applications across six distinct use cases currently in development, each entering an initial validation phase of a two-cycle process. The goal is to collect data through subjective UX assessment and observations. Results from these preliminary user tests will inform the next iteration in a user-centered development approach. The applications span VR, AR (mobile and head-mounted), and desktop platforms, covering the following use cases: DT Creation, City Planning, City Maintenance, VR Virtual Tourism, AR Virtual Tourism, AR Industrial DT Maintenance. In each of these use cases, AI plays a crucial role in the simulation, reconstruction, and recognition of the DT environments. For example, AI algorithms are used to model real-world behaviors and dynamics (traffic patterns, seasonal changes), while advanced 3D reconstruction techniques are used to create realistic digital environments, increasing the user's immersion. Furthermore, accurate image and object recognition enable seamless interaction by guiding users in real-time.

%\begin{itemize}
%    \item DT Creation
%    \item City Planning
%    \item City Maintenance
%    \item VR Virtual Tourism
%    \item AR Virtual Tourism
%    \item AR Industrial DT Maintenance
%\end{itemize}

The evaluation of applications will follow an on-field approach for the AR use cases and in-lab testing for desktop and VR applications, using a Minimum Viable Product (MVP) version of each application. Users are given concrete tasks to complete but are also encouraged to freely explore and interact with the applications. Direct observations will be gathered to understand how users engage with the MVPs and to capture their immediate reactions. At the end of the experience, the participants will complete a mix of standardized and custom-tailored questionnaires designed to assess their subjective impressions regarding the experiences and to measure different usability metrics such as cognitive load.

This combination of observational and subjective evaluations aims to provide a comprehensive understanding of each application's effectiveness and UX to identify areas for improvement, supporting the ongoing development of the DT applications.

\subsection{Applications for the first validation phase}
The functionalities of the MVPs for each use case and the tasks planned for the user studies are briefly summarized in the following:

\subsubsection{DT Creation} The MVP to be tested for the DT Creation use case is a desktop application that enables users to create and customize a DT from point cloud data. They can explore the city in a 3D editor view and then create the DT with a single mouse click, that transforms the point cloud into meshes. It also facilitates asset management, allowing users to replace low-quality objects with high-quality versions via a drag-and-drop from an asset library interface. The task required from the participants will be the creation of the DT, followed by an exemplary exchange of a building through the UI.

\subsubsection{VR Virtual Tourism}
The MVP for the VR Virtual Tourism use case is a standalone app for Meta Quest, based on the DT created in the DT Creation application. In this VR version, users can explore the city’s DT through the head-mounted display (HMD) in an immersive first-person mode, either through continuous movement or by teleporting between Points of Interest (POIs). Users can access an interface with weather parameters, allowing them to adjust the sun's intensity and enable rain. Additionally, they can take snapshots of the virtual environment, similar to taking photos in the physical world. During the study, participants will be asked to try both movement forms, change the weather parameters, and take snapshots of city landmarks.

\begin{figure}[ht]
    \centering
    \includegraphics[width=\linewidth]{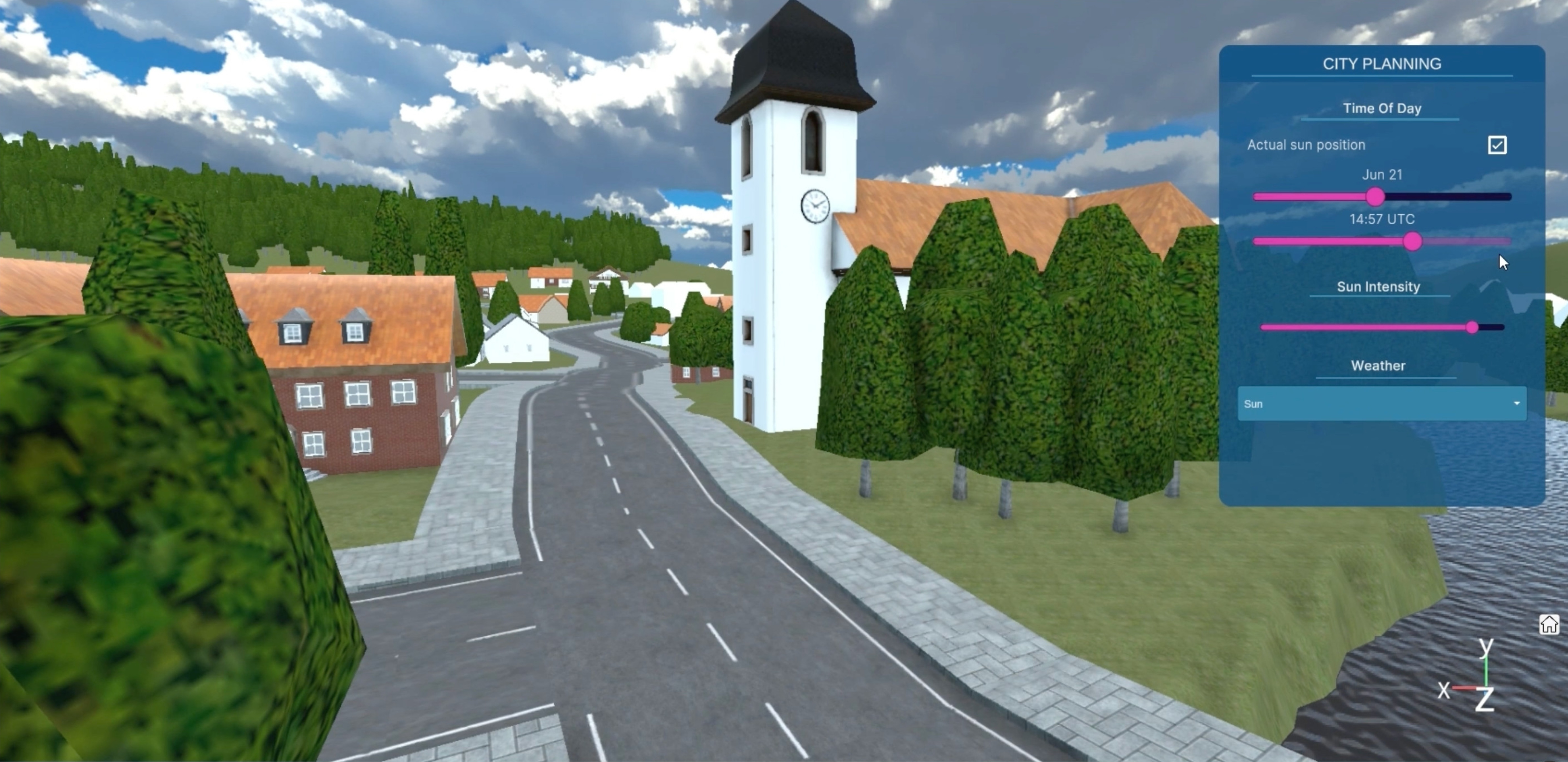}
    \caption{Screenshot of the MVP for the City Planning use case that shows a POI (church) together with the weather UI.}
    \label{fig:CP}
\end{figure}

\subsubsection{City Planning}
The City Planning MVP is also based on the DT Creation MVP. In this desktop application, users start with a 3D overview of the DT and can navigate using a mouse or trackpad. They can access a weather interface to adjust sun intensity and simulate rain, as seen in Figure \ref{fig:CP}. Additionally, they can change the time of day to observe the sun's accurate positioning. For the user test, participants will be required to navigate the DT and adjust specific weather parameters.

\subsubsection{AR Virtual Tourism} 
\begin{figure}[ht]
    \centering
    \includegraphics[width=\linewidth]{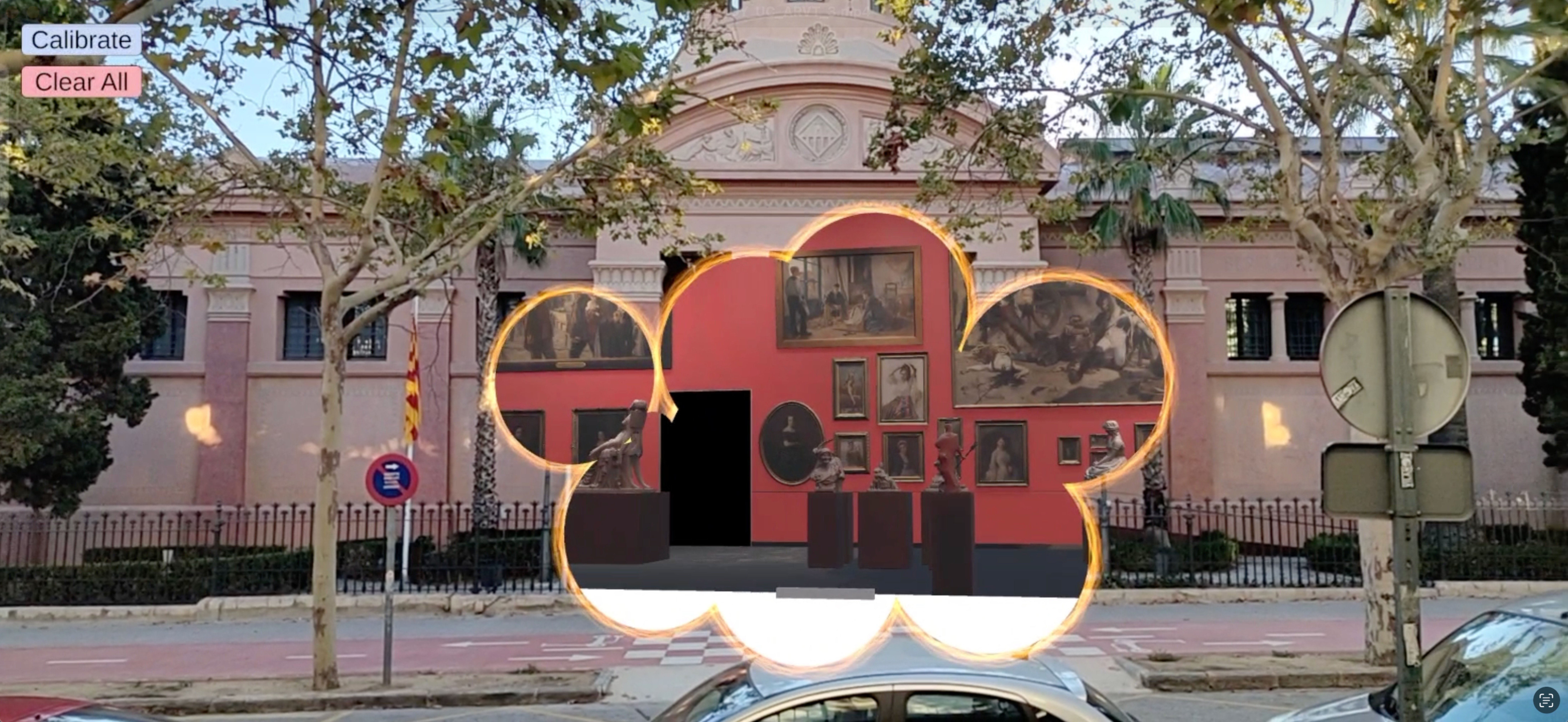}
    \caption{Screenshot of the MVP for the AR Virtual Tourism use case that shows an open portal in front of a POI (museum).}
    \label{fig:ARVT}
\end{figure}

The MVP for the AR Virtual Tourism use case is a mobile application. For testing, participants are located on-site in the city in front of a pre-selected building and are provided with a mobile device. The user can place a virtual AR portal in the scene by tapping on the desired position on the phone's screen. The portal opens, allowing the user to peek inside the building, displaying the interior of the real-world structure, as visualized in Figure \ref{fig:ARVT}. Additionally, information about the POI can be accessed via a tap. During the study, participants will be asked to first create a portal, step inside the virtually placed interior, and lastly access the building's information.
\subsubsection{City Maintenance}
For the City Maintenance use case, a web-based application was developed to be used on a desktop PC in the initial validation phase. Through a top-down map overview of the DT area, users can access maintenance issue reports processed from pre-recorded data curated for user testing. The reports must first be reviewed and confirmed by the user to rule out classification errors. Additionally, users can rate each issue and assign it a quality score. For user testing, participants will be required to verify the accuracy of several issue reports and provide a quality rating

\subsubsection{AR Industry DT Maintenance}
The MVP for the AR Industry DT Maintenance application is designed for HoloLens. Users are located in a real-world warehouse wearing the HoloLens 2. Through the headset, an AR overlay of the DT is displayed. If users detect a mismatch between the physical world and the DT, they can realign the DT or rescan the environment if necessary. Additionally, users can flag POIs, such as charging stations for autonomous robots. During the study, participants will be asked to perform realignment, rescanning, and POI labeling, resulting in three distinct tasks.

\subsection{Questionnaires}
Standardized questionnaires will be used to assess technological affinity, task load, emotional response, presence, cybersickness, and UX. The selected questionnaires are widely used in UX and XR research and it is expected that their usefulness holds true for applications that use DTs. In the future it could be valuable to work towards standardized questionnaires specifically developed to measure the UX with DTs. Custom questions were developed to explore specific aspects of the application laying the basis for further iteration of the use cases. The choice of questionnaires is consistent across use cases, except for VR Virtual Tourism. For the AR and desktop applications, Demographics, ATI, NASA-TLX, SAM, UEQ-S and a custom questionnaire will be included. In contrast, the testing of the VR use case will include the IPQ and CSQ-VR as this use case focuses on immersion and there is a risk of cybersickness. The NASA-TLX won't be used here, as the explorative nature of the experience reduces the emphasis on completing specific tasks.

\subsubsection{Demographics} 
The demographics questionnaire collects general information, including age, gender, profession, and prior experience with XR technologies.

\subsubsection{ATI} 
The Affinity for Technology Interaction (ATI) score \cite{franke2019personal} can be calculated from a 9-item questionnaire to assess an individual’s tendency to actively engage with technology. It can help observe trends potentially related to overall experience with technology.

\subsubsection{SAM}
The Self-Assessment Manikin (SAM) \cite{bradley1994measuring} assesses three emotional dimensions: Valence, Arousal, and Dominance. Each dimension includes five pictorial representations of emotions with increasing intensity.
\subsubsection{IPQ} 
The iGroup Presence Questionnaire (IPQ) \cite{igroup} is a widely used tool that distinguishes between Spatial Presence, Involvement, Realness, and Overall Presence making it relevant for immersive experiences. 

\subsubsection{NASA-TLX}
The NASA Task Load Index (NASA-TLX) \cite{Hart_Staveland_1988} is a commonly used tool for assessing cognitive and physical demands, among others.

%It features six question items, each originally designed with 7-point scales that included three levels of intensity, resulting in a total of 21 increments. In this study, only the 7-point scale will be used due to the limitations of the online questionnaire. 

\subsubsection{UEQ-S}
The short version of the User Experience Questionnaire (UEQ-S) \cite{Laugwitz_Held_Schrepp_2008} consists of eight scales, each representing a pair of attributes. It is used to assess users' perceptions of both the pragmatic and hedonic quality of a system or application.
%The UEQ-S will be used for testing in all use cases.

\subsubsection{CSQ-VR} 
The Cybersickness in Virtual Reality Questionnaire (CSQ-VR) \cite{CSQ-VR}\cite{ Kourtesis_Linnell_Amir_Argelaguet_E_MacPherson_2023} was chosen to assess the degree of cybersickness caused by the use of VR, over the more commonly used Simulator Sickness Questionnaire (SSQ) due to criticism in recent years \cite{Bimberg_Weissker_Kulik_2020}. The CSQ-VR is an alternative that has been validated against the SSQ. It features two question items for each category (Nausea, Vestibular, and Oculomotor). Additionally, participants can provide qualitative feedback if further specification of symptoms is needed. 
%The CSQ-VR will be used in the VR Virtual Tourism use case, as it is specifically relevant for applications using HMDs.

\subsubsection{Custom questionnaires} 
Each user study concluded with a final non-standardized questionnaire regarding specific aspects of the applications. Participants will be asked to rate their overall satisfaction with the application (\enquote{How much did you enjoy the experience?}) and the ease of use and usefulness of the implemented features (e.g.,\enquote{How easy was it to find the point of interest?},\enquote{How would you rate the usefulness of this feature?}). Lastly, participants will be provided an input field to offer space for additional feedback and comments. The results from the custom questionnaires will help guiding the next iteration of the prototypes, forming the foundation for refining and enhancing the tested features.

\subsection{Participants}
The recruitment goal is to enlist a minimum of 20 participants for each of the two locations to ensure adequate statistical power for evaluating subjective data. Participants will be recruited directly from the locations where user testing will be conducted, including a mix of external volunteers and employees of the institutions involved. It will be ensured that these participants are neither familiar with nor associated with the project to prevent bias. Additionally, stakeholders are planned to be involved in the validation process, through remote testing offered for use cases with desktop applications.

\section{Conclusion}
User-centric evaluation of XR-enhanced DTs is essential for improving interaction quality and optimizing usability across diverse applications. Through a multi-faceted evaluation approach—incorporating real-world testing, task-based usability assessments, and targeted questionnaires this research provides a comprehensive methodology for analyzing UX, ease of interaction, and cognitive load. Moreover, this structured approach to user testing and feedback collection establishes benchmarks for future DT-XR projects, supporting industry-wide standards for usability, cognitive load management, and immersive experience quality. A validation with more than 40 users in two cities will be carried out with the described protocol to assess the usability of the presented use case on the field evaluations. The results of the employed questionnaires will be evaluated to refine the validation methodology. This will involve examining the insights provided by standardized UX research questionnaires and identifying areas where specific DT-focused questionnaires could be beneficial.
The evaluation methodologies presented in this research also provide a foundation for refining AI components through user feedback and creating DTs with better simulation, reconstruction, and recognition capabilities based on real-world usability data.
\section*{Acknowledgment}
This work was supported by the European Union's Horizon Europe programme under grant number 101092875 ``DIDYMOS-XR'' (https://www.didymos-xr.eu).

The authors used ChatGPT-4 to review and correct parts of the English text to ensure clarity, accuracy, and grammatical consistency throughout the document.

\bibliographystyle{IEEEtran}
\bibliography{main}

\end{document}